\documentclass{article}
\usepackage{amsmath}
\usepackage[nointegrals]{wasysym}
\usepackage{makecell}
\usepackage{array}
\usepackage{mathrsfs}
\usepackage{mathtools}
\usepackage{amssymb}
\usepackage{graphicx}
\usepackage{float}
\usepackage{authblk}
\usepackage{cite}
\usepackage[symbol]{footmisc}
\usepackage[left=3cm, right=2cm, top=2cm, bottom=2cm]{geometry}
\usepackage{hyperref}
\usepackage{fancyhdr}
\graphicspath{{fig/}}
\usepackage{caption}
\begin{document}
\title{\textbf{Identifying diverging-effective mass in 
MOSFET and $^3$He systems}\footnote{This was presented and accepted in ICEIC 2022 Conference held on 6th$-$9th of February in 2022 at Grand Hyatt Jeju in Republic of Korea.}}

\author {Hyun-Tak Kim{\footnote{\href{mailto:htkim@etri.re.kr}{htkim@etri.re.kr, htkim0711@snu.ac.kr, htkim580711@gmail.com}}}} 

\font\myfont=cmr12 at 10pt
\affil{\myfont Metal-Insulator Transition Lab., Electronics and Telecommunications Research Institute, Daejeon 34129, South Korea}

\date{}

\maketitle

\vspace*{-1cm}

\begin{abstract}
\font\myfont=cmr12 at 11pt
\noindent{\myfont Emerging devices such as a neuromorphic device and a qubit can use the Mott transition phenomenon, but in particular, the diverging mechanism of the phenomenon remains to be clarified.  The diverging-effective mass near Mott insulators was measured in strongly correlated Mott systems such as a fermion $^3$He and a Si metal-oxide-semiconductor-field-effect transistor, and is closely fitted by the effective mass obtained by the extension of the Brinkman-Rice(BR) picture, $m^*/m=1/[1-(U/U_c)^2]=1/(1-{\kappa}^2_{BR}{\rho}^4)$ when ${\kappa}^2_{BR}{\approx}1({\neq}$1), where $0<U/U_c={\kappa}_{BR}{\rho}^2<1$, correlation strength is ${\kappa}_{BR}$, band-filling is ${\rho}$. Its identification is a percolation of a constant mass in the Brinkman-Rice picture. Over ${\kappa}_{BR}{\approx}0.96$ is evaluated.
}

\end{abstract}


\vspace*{0.5cm}
\font\myfont=cmr12 at 12pt
\noindent\textbf{I. INTRODUCTION}\\
\indent The diverging-effective mass (DEM) in a strongly correlated metallic Mott system \cite{brinkman1970application} is evidence of strong correlation between fermions that often occurs in condensed phenomena such as superconductivity, superfluid, the Mott transition, and a fractional quantum Hall effect. The characteristics of the DEM can be very important to reveal the mechanism of the phenomena. 
Although the DEM has been measured in materials such as Sr$_{1-x}$La$_x$TiO$_3$ \cite{tokura1993LSTO}, VO$_2$ \cite{qazilbash2007Mott} a 2-dimensional (dim) fermion $^3$He \cite{casy2003He3,neumann2007He3}, a dilute 2-dim metal-oxide-semiconductor-field-effect transistor (MOSFET)\cite{mokashi2012MOSFET}, and a superconductor YBa$_2$Cu$_3$O$_{6+x}$ \cite{sebastian2010super}, the identification of the DEMs still remains to be revealed \cite{spivak2010colloquim}.\\

\noindent\textbf{II. PROPOSED METHOD}\\
\indent Here, the metal-insulator transition (MIT) theory with the DEM is briefly reviewed for less than and one electron per atom. The DEM, the inverse of discontinuity,$d$, at $E_F$ [1], means the increase of the number of carriers outside $E_F$,
 
\begin{equation}
\begin{split}
\frac{1}{d}\equiv\frac{m^*}{m}=\frac{1}{[1-(U/U_c)^2]},
\\=\frac{1}{(1-{\kappa}^2_{BR}{\rho}^4)}.
\label{1}
\end{split}
\end{equation}

\indent This can be satisfied when the Coulomb-repulsion ratio 0$<U/U_c={\kappa}_{BR}{\rho}^2<$1, band filling $0<{\rho}=n_{metal}/n_{tot}{\leq}1$, correlation strength 0$<U/U_c={\kappa}{\leq}$1 when ${\rho}$=1 half filling. 
In these expressions, $n_{metal}$ is the carrier density in the metal region, $n_{tot}$ is the number of all atomic sites in the 
measurement region, and $U_c$ is the critical Coulomb interaction \cite{brinkman1970application}\cite{kim2000extension,kim2001extended}. 
Eq. (1) was derived by an extension of the Brinkman-Rice (BR) picture using a fractional effective charge (or fractron), $e^*={\rho}e$, averaged over all atomic sites in the measurement region \cite{kim2000extension,kim2001extended,kim2007mobrik}. The Mott insulator at $U/U_c$=1 is assumed when ${\rho}=$1. Existence of a DEM in a hole-doping region where $n=n_{tot}-n_c$ (about 0.02$\%$ from Mott criterion \cite{kim2004mott}) can be seen as a hole-driven Mott MIT. 
Eq. (1) gives a measurable quantity that represents the average of the true $m^*/m$ in the BR picture. Moreover, the BR picture \cite{brinkman1970application} was also confirmed in the metallic side of $d={\infty}$ Hubbard model \cite{zhang1993hubbard}.
The DEM was observed in $^3$He at high density \cite{casy2003He3}, which is a prototype system for the strongly correlated system. This is explained in Figs. 1A and 1B, which shows that the quantity ${\rho}=n/n_{tot}$ increases as the $^3$He density increases (where $n_{tot}{\approx}5.101 nm^{-2}$). The increasing ${\rho}$ induces a DEM, as per Eq. (1), which closely fits the experimental data (Fig. 1B). 
Another example helps clarify the diverging thermoelectric power measured in Si-MOSFET systems \cite{mokashi2012MOSFET}. As the carrier density decreases, ${\rho}=n_c/n_s$ increases (Fig. 2C). This is why the extent of metal region is constant (about 0.02$\%$) - because a conducting path with lower resistance is formed in the channel of MOSFET after the insulator-to-metal transition occurs. The DEM data for this system \cite{casy2003He3} are closely fitted by Eq. (1) (Fig. 2D).\\

\noindent\textbf{III. CONCLUSIONS}\\
\indent In conclusion, the DEM identified in many Mott systems is the effect of measurement due to inhomogeneity and percolation phenomenon. The true value of the DEM in such systems is the renormalized constant mass in the BR picture\cite{brinkman1970application}. The correlation strengths in 3He and MOSFET systems associated with the DEM are ${\kappa}_{BR}{\approx}$0.96 ($m^*/m{\approx}$13) determined from $m^*/m=1/(1-{\kappa}_{BR}^2)$ in assuming ${\rho}_c{\equiv}{n/n_c}{\approx}$0.98 when ${\rho}{\approx}$1. This reveals that the $^3$He system and the MOSFET system are strongly correlated. Diverging in thermoelectric power is also attributed to the DEM. The Mott transition near $\rho$=1 can be used for future emerging devices such as a neuromorphic device and a qubit.

\begin{figure}[H]
\centering
\includegraphics [scale=0.4]{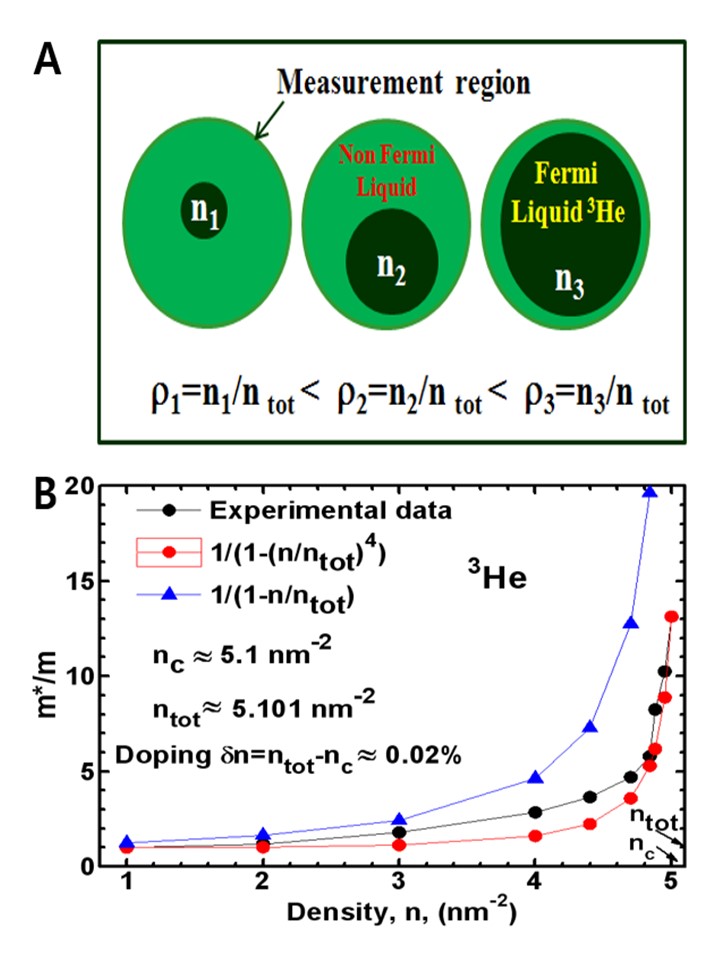}
\font\myfont=cmr12 at 11pt
\caption {\myfont (A) Schematic diagram understanding the Fermi-liquid region in a $^3$He (hole doping) system with monolayer. $n_i=n_{Fermi}$ and $n_{non-Fermi}$ are particle densities in black and green regions, respectively, and $n_{tot}=n_{Fermi}+n_{Non-Fermi}$. Band filling, ${\rho}_i=n_i/n_{tot}$, is given. As $n_i$ increases, ${\rho}_i$ increases; this is a percolation phenomenon. (B) Experimental data \cite{casy2003He3} and fittings using Eq. (1) when ${\kappa}_{BR}{\approx}1({\neq}1)$ are shown.}
\end{figure}

\begin{figure}[H]
\centering
\includegraphics [scale=0.6]{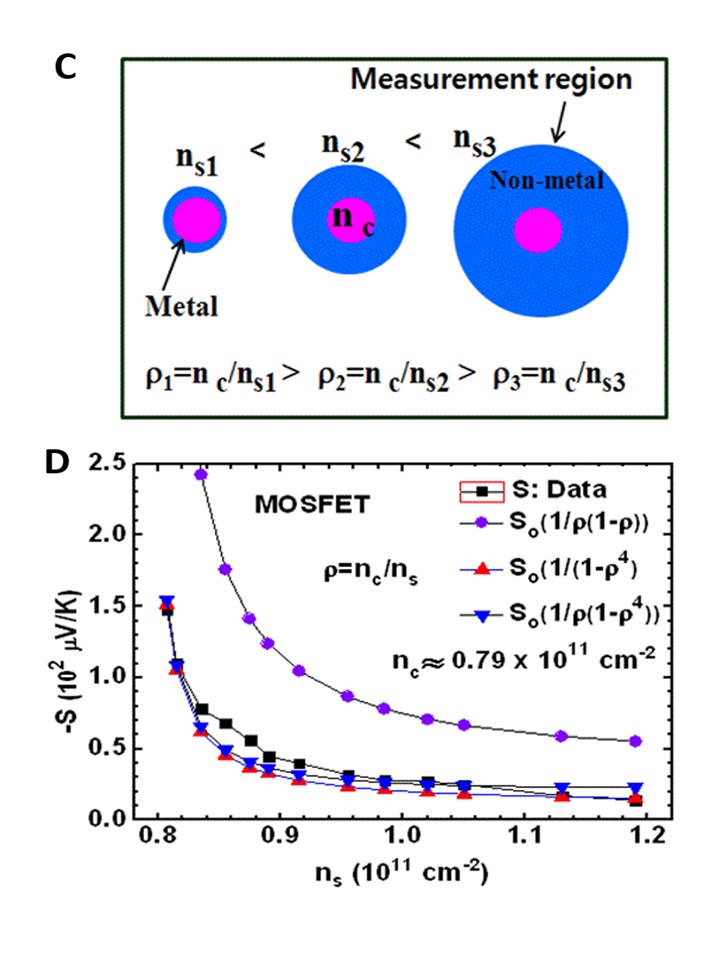}
\font\myfont=cmr12 at 11pt
\caption {\myfont (C) Schematic diagram of the metal region in a Si-MOSFET (electron doping). Although the carrier density, $n_s$, increases, the extent of the metal phase (pink region) with $n_c{\approx}0.79 $x$ 10^{-11} cm^{-2}$ is constant (about 0.02$\%$ to $n_{Si}{\approx}3.4 $x$ 10^{-14} cm^{-2}$ in Si), but $n_{non}$ in the non-metal region (blue region) increases (the increase of the measurement region); $n_{tot}{\equiv}n_s=n_c+n_{non}$. Given ${\rho}=n_c/n_s$ increases as $n_s$ decreases. (D) Experimental data extracted from the thermoelectric power measured at 0.8 K \cite{mokashi2012MOSFET} is closely fitted by Eq. (1). The thermoelectric power of metal, $S=({\alpha}{\pi}^3k^2_BT/3e)(1/E_F) =(8{\alpha}{\pi}^3k^2_BT/3h^2)(m^*/e^*n_c) =S_o(1/{\rho})(1/(1-{\rho}^4))$, is given, where $e^*={\rho}e$  \cite{kim2000extension,kim2001extended,kim2007mobrik}, ${\rho}=n_c/n_s$, $T=$0.8 K, $m^*=m_o/(1-{\rho}^4), {\alpha}=$0.6, and $S_o=(8{\alpha}{\pi}^3k^2_BT/3h^2)(m_o/en_c)\approx$20.6 are used.}
\end{figure}

\renewcommand\theadalign{bc}
\renewcommand\theadfont{\bfseries}
\renewcommand\theadgape{\Gape[4pt]}
\renewcommand\cellgape{\Gape[4pt]}

\noindent \textbf{ACKNOWLEDGEMENTS}\\
\noindent \small{This work was supported by Institute of Information and Communications Technology Planning and Evaluation (IITP) grant funded by the Korean government (MSIT) via Grant 2017-0-00830. I acknowledge Y. S. Lee, T. Nishio, T. Driscoll, S. Kravchenko for valuable comments. Parts of this research were presented in 2013 APS March Meeting, Abstract: U19.00007 : Fitting of Diverging Thermoelectric Power in a Strongly Interacting 2D Electron System of Si-MOSFETs,\href{https://meetings.aps.org/Meeting/MAR13/Session/U19.7}, and 2012 APS March Meeting, Abstract: A16.00015 : Fitting of m*/m with Divergence Curve for $^3$He   Fluid Monolayer using Hole-driven Mott Transition,\href{https://meetings.aps.org/Meeting/MAR12/Event/158850}.

\end{document}